\RequirePackage[mathlines]{lineno} 
\documentclass[a4paper,11pt]{article}

\usepackage{jheppub} 

\usepackage[T1]{fontenc} 

\usepackage{threeparttable}
\usepackage{multirow}
\usepackage{subfigure}
\usepackage{float}
\usepackage{ulem}
\usepackage{caption}

\let\emph\textit
\let\oldequation\equation
\let\oldendequation\endequation
\renewenvironment{equation}
  {\linenomathNonumbers\oldequation}
  {\oldendequation\endlinenomath}

\def \ee   {e^+e^-}
\def \jpsi   {J/\psi}

\def \pp   {\pi^+\pi^-}
\def \gev  {\mbox{GeV}}

\begin{document}

\title{\boldmath Measurement of the cross section of $e^{+}e^{-}\to\eta\pi^{+}\pi^{-}$ at center-of-mass energies from 3.872 GeV to 4.700 GeV}
\author{\small
M.~Ablikim$^{1}$, M.~N.~Achasov$^{10,b}$, P.~Adlarson$^{68}$, S. ~Ahmed$^{14}$, M.~Albrecht$^{4}$, R.~Aliberti$^{28}$, A.~Amoroso$^{67A,67C}$, M.~R.~An$^{32}$, Q.~An$^{64,50}$, X.~H.~Bai$^{58}$
% ...
\\
\vspace{0.2cm}
(BESIII Collaboration)\\
\vspace{0.2cm} {\it
$^{1}$ Institute of High Energy Physics, Beijing 100049, People's Republic of China\\
% ...
$^{4}$ Bochum Ruhr-University, D-44780 Bochum, Germany\\
% ...
$^{10}$ G.I. Budker Institute of Nuclear Physics SB RAS (BINP), Novosibirsk 630090, Russia\\
% ...
$^{14}$ Helmholtz Institute Mainz, Staudinger Weg 18, D-55099 Mainz, Germany\\
% ...
$^{28}$ Johannes Gutenberg University of Mainz, Johann-Joachim-Becher-Weg 45, D-55099 Mainz, Germany\\
% ...
$^{32}$ Liaoning Normal University, Dalian 116029, People's Republic of China\\
% ...
$^{50}$ State Key Laboratory of Particle Detection and Electronics, Beijing 100049, Hefei 230026, People's Republic of China\\
% ...
$^{58}$ University of Jinan, Jinan 250022, People's Republic of China\\
% ...
$^{64}$ University of Science and Technology of China, Hefei 230026, People's Republic of China\\
% ...
$^{67}$ University of Turin and INFN, (A)University of Turin, I-10125, Turin, Italy; (B)University of Eastern Piedmont, I-15121, Alessandria, Italy; (C)INFN, I-10125, Turin, Italy\\
$^{68}$ Uppsala University, Box 516, SE-75120 Uppsala, Sweden\\
% ...
$^{b}$ Also at the Novosibirsk State University, Novosibirsk, 630090, Russia\\
% ...
}\vspace{0.4cm}}

\abstract{
Using data samples with an integrated luminosity of 19 fb$^{-1}$ at twenty-eight center-of-mass energies from 3.872 GeV to 4.700 GeV collected with the BESIII detector at the BEPCII electron--positron collider, the process $e^{+}e^{-}\to\eta\pi^{+}\pi^{-}$ and the intermediate process $e^{+}e^{-}\to\eta\rho^{0}$ are studied for the first time. The Born cross sections are measured. No significant resonance structure is observed in the cross section lineshape.}

\maketitle
\flushbottom

\section{\boldmath Introduction}
\label{sec:intro}
Since the discovery of the charmonium-like state $X$(3872) by
Belle~\cite{x3872}, a series of states, such as $Y$(4260)~\cite{y4260}
and $Z_{c}$(3900)~\cite{zc1,zc2}, which are unexpected in charmonium
spectroscopy, have been found. The $Y$(4260) was observed by the
BaBar Collaboration in
$e^{+}e^{-}\to\gamma_{\rm{ISR}}\pi^{+}\pi^{-}J/\psi$, where the
subscript ISR stands for initial state radiation, and confirmed by
CLEO and Belle~\cite{intro4,intro5}. In 2017, the BESIII Collaboration
performed a dedicated scan of the process $e^{+}e^{-}\to~\pi^{+}\pi^{-}J/\psi$ at
center-of-mass energies $\sqrt{s}$ from 3.77 to 4.60 GeV. Two structures were
observed with masses of $M = 4222.0 \pm 3.1 \pm 1.4$ MeV/$c^{2}$ and
$M = 4320.0 \pm 10.4 \pm 7.0$ MeV/$c^{2}$~\cite{intro6}. The former
one was regarded as the previously observed $Y$(4260), which was renamed
to $Y$(4220). The $Y$(4220) state was confirmed by Born cross section
measurements of the final states $\omega\chi_{c0}$~\cite{intro7},
$\pi^{+}\pi^{-}h_{c}$~\cite{intro8},
$\pi^{+}\pi^{-}\psi(3686)$~\cite{intro9}, and
$\pi^{+}D^{0}D^{*-}$~\cite{intro10} by BESIII.

The currently known decays of $Y$(4220) occur only to open or
hidden-charm final states. However, some related theories point out
that charmonium-like states are also likely to decay to light hadron
final states~\cite{light1}, shedding further light on the $Y$(4220)~\cite{intro11}. Several measurements of the cross
sections for $e^+e^-$ annihilations to light hadrons have been measured by the
BESIII Collaboration, such as $e^{+}e^{-}\to
K_{S}^{0}K^{\pm}\pi^{\mp}\pi^{0}$~\cite{light2},
$K_{S}^{0}K^{\pm}\pi^{\mp}\eta$~\cite{light3}, $p\bar{n}K^{0}_{S}K^{-}
+ c.c.$~\cite{light4}, $p\bar{p}\pi^{0}$~\cite{light5} etc., but no
significant structures have been found so far.  In order to better
understand the composition and properties of the charmonium-like
states, further searches for their decays to
charmless light-hadron final states are important.

In this paper, we present the measurements of the Born cross section
of the process $e^{+}e^{-}\to\eta\pi^{+}\pi^{-}$ at center-of-mass
(c.m.) energies from 3.872 GeV to 4.700 GeV, and search for possible
charmonium ($\psi$) or vector charmonium-like ($Y$) states in the
corresponding lineshape.

\section{BESIII detector and data sets}
The BESIII detector is a magnetic spectrometer~\cite{BESIII} located at the Beijing Electron Positron Collider (BEPCII). The cylindrical core of the BESIII detector consists of a helium-based multilayer drift chamber (MDC), a plastic scintillator time-of-flight system (TOF), and a CsI (Tl) electromagnetic calorimeter (EMC), which are all enclosed in a superconducting solenoidal magnet providing a 1.0 T magnetic field. The solenoid is supported by an octagonal flux-return yoke with resistive plate counter muon identifier modules interleaved with steel. The acceptance of charged particles and photons is 93\% over 4$\pi$ solid angle. The charged-particle momentum resolution at 1 GeV/$c$ is 0.5\%, and the specific ionization energy loss ($dE/dx$) resolution is 6\% for the electrons from Bhabha scattering. The EMC measures photon energies with a resolution of 2.5\% (5\%) at 1 GeV in the barrel (end cap) region. The time resolution of the TOF barrel part is 68 ps, while that of the end cap part is 110 ps. The end cap TOF system was upgraded in 2015 with multi-gap resistive plate chamber technology, providing a time resolution of 60 ps~\cite{BESTOF1,BESTOF2}.

The twenty-eight data sets taken at $\sqrt{s}=3.872\sim4.700~\gev$ are
used in this analysis. The nominal energy of each data set is
calibrated by the process
$e^{+}e^{-}\to(\gamma_{\rm{ISR/FSR}})\mu^{+}\mu^{-}$~\cite{CMS1,CMS2},
where the subscript $\rm{FSR}$ stands for final-state radiation. The
integrated luminosity $\mathcal{L}$ is determined by large angle
Bhabha events~\cite{Lum1,Lum2}, and the total integrated luminosity is
approximately 19 fb$^{-1}$.

\section{Monte Carlo (MC) simulation}
Simulated data samples produced with {\sc geant4}-based~\cite{BESIII5}
MC software, which includes the geometric description of the BESIII
detector and the detector response, are used to determine detection
efficiencies and to estimate backgrounds. The simulation models the
beam energy spread and ISR in the $\ee$ annihilations with the
generator {\sc kkmc}~\cite{BESIII6}. The inclusive MC simulation
sample includes the production of open charm processes, the ISR
production of vector charmonium(-like) states, and the continuum
processes incorporated in {\sc kkmc}. The known decay modes are
modeled with {\sc evtgen}~\cite{BESIII7} using branching fractions
taken from the Particle Data Group (PDG)~\cite{BESIII8}, and the
remaining unknown $\psi$ decays are modeled with {\sc lundcharm}~\cite{BESIII9}. The FSR from charged final state
particles is incorporated using {\sc photos}~\cite{BESIII10}.

In the signal MC simulation samples at each c.m.\ energy point, three
exclusive processes are involved, which are the three-body
non-resonant process $\ee\to\eta\pp$, and the two-body resonant
processes, $\ee\to a_{2}^{\pm}(1320)\pi^{\mp}$ and $\eta\rho^{0}$. The
last process is simulated by the HELAMP model~\cite{BESIII7} following the dynamics of other vector charmonium decays, while the
other two processes are simulated by phase space (PHSP) models. In
determining the resulting detection efficiencies, the interference
between $\eta\rho$ and the three-body non-resonant processes is
included, while the interference between $a_{2}(1320)\pi$ and the
three-body non-resonant process is neglected due to the low
statistics. The resulting detection efficiency is obtained by mixing
the three processes weighted according to the number of
observed events ($N_{\rm obs}$) and detection efficiency ($\epsilon$).

\section{Event selection}
The charged tracks detected in the MDC are required to be within a
polar angle ($\theta$) range of $|\textup{cos} \theta| < 0.93$, where
$\theta$ is defined with respect to the $z$-axis, which is the
symmetry axis of the MDC. All the charged tracks are required to
originate from the interaction region $V_{xy} <$ 1 cm and $|V_{z}| <$
10 cm, where $V_{xy}$ and $|V_{z}|$ are the distances of closest
approach of the charged track to the interaction point in the
$xy$-plane and $z$ direction, respectively.

Particle identification (PID) for charged tracks combines measurements
of $dE/dx$ in the MDC and the flight time in the TOF to form
likelihoods $\mathcal{L}(h)$ for each hadron $h=p,K,\pi$
hypothesis. Tracks are identified as pions when the pion hypothesis
has the greatest likelihood ($\mathcal{L}(\pi)>\mathcal{L}(K)$ and
$\mathcal{L}(\pi)>\mathcal{L}(p)$).

Photon candidates are identified using showers in the EMC. The
deposited energy of each shower must be more than 25 MeV in the barrel
region ($|\textup{cos} \theta| <$ 0.80) and more than 50 MeV in the
end cap region (0.86 $ < |\textup{cos} \theta| <$ 0.92). To exclude
showers that originate from charged tracks, the angle between the
position of each shower in the EMC and the closest extrapolated
charged track must be greater than 10 degrees. To suppress electronic
noise and showers unrelated to the event, the difference between the
EMC time and the event start time is required to be within [0,700]
ns. Candidate events must have two charged tracks with zero net
charge, and the number of photons should be 2 or greater. The two
charged tracks must be identified as pions.

To improve the momentum and energy resolution and suppress the
potential backgrounds, a four-constraint (4C) kinematic fit, which
constrains the total four-momentum of the final state particles to
that of the initial colliding beams, is applied to the event under the
hypothesis of $\ee\to\gamma\gamma\pp$. If more than one candidate
(37\% of the selected events) exists in an event, that with the smallest
$\chi^{2}_{4C}$ is selected.

Potential backgrounds are investigated with six equivalent-luminosity
inclusive MC samples generated at c.m. energies from 4.009 GeV to
4.600 GeV, using an event-type analysis tool,
TopoAna~\cite{TopoAna}. It is found that the main background
contributions come from $e^{+}e^{-}\to\gamma\pi^{+}\pi^{-}$,
$\mu^{+}\mu^{-}, \gamma\gamma e^{+}e^{-}$ and $e^{+}e^{-}\to
\jpsi+{\rm anything},\jpsi\to hh~(h=p,\pi,e,\mu)$ processes. In the
first background channel, a reconstructed photon \emph{e.g.} from beam-related background
is combined with the real photon to form a fake $\eta$ signal. This
background is suppressed by requiring the ratio $R=\frac{\mid
E_{\gamma_1} - E_{\gamma_2}\mid}{p_{\eta}} < 0.90$, where
$E_{\gamma_1}, E_{\gamma_2}$ are the energies of the two photons from the
fake $\eta$ decay and $p_{\eta}$ is the momentum of the fake $\eta$.
The second background channel is suppressed by requiring the hit depth
of charged track in the $\mu$ counter to be less than 40 cm. The third
background channel is suppressed by requiring $E/cp<$0.7. Here, $E$ and
$p$ denote the deposited energy in the EMC and the momentum of the charged
track, respectively. The background from $\jpsi$-related events is
vetoed by requiring the invariant mass of $\pp$ not to fall into the
$\jpsi$ mass region [3.05,~3.15] GeV/$c^{2}$. Finally, it is found
that the dominant remaining background channel is
$e^{+}e^{-}\to\mu^{+}\mu^{-}$ due to $\mu-\pi$ misidentification.

\begin{figure}[htbp] \centering
\includegraphics[width=0.47\textwidth]{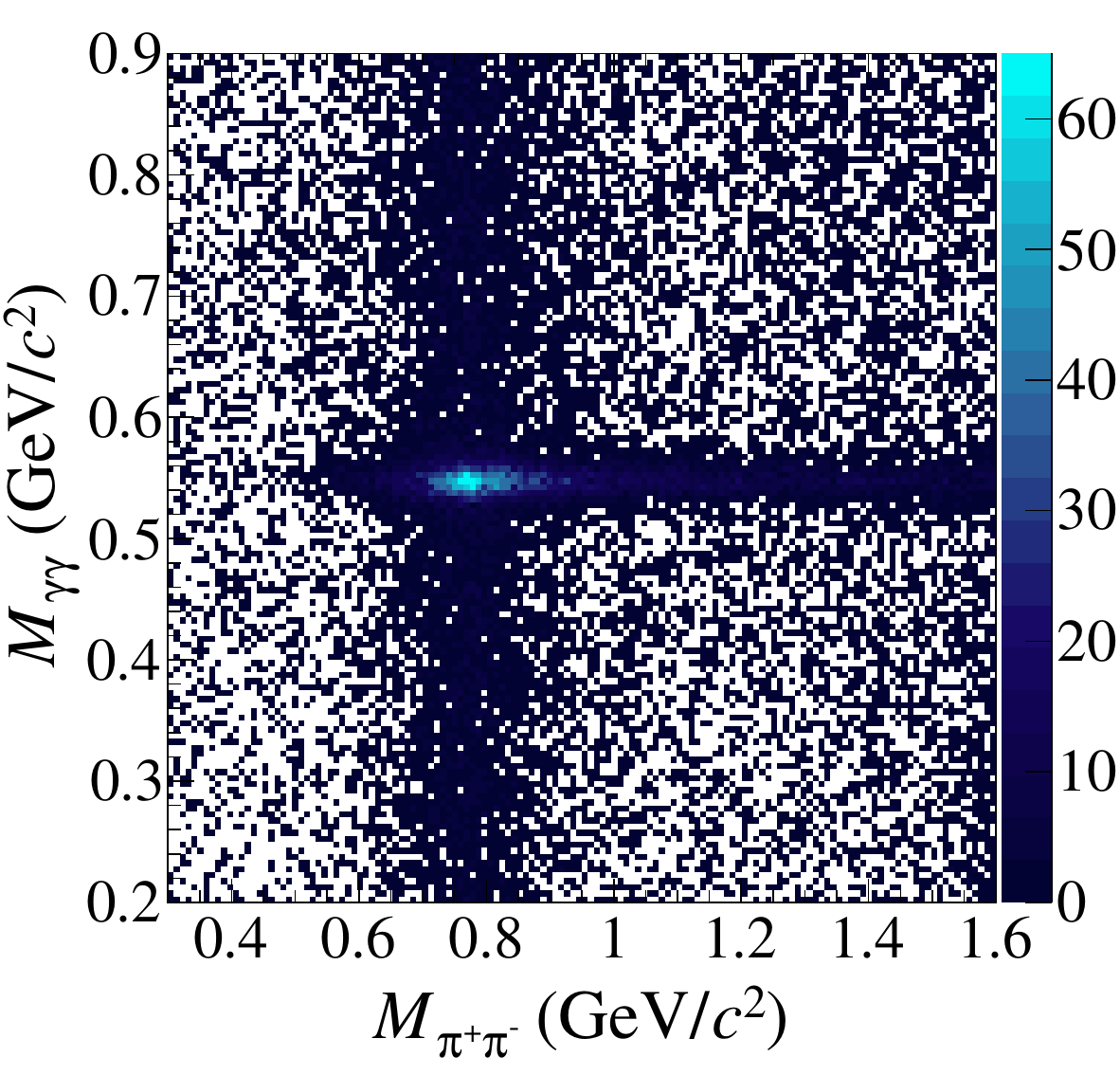}
\includegraphics[width=0.47\textwidth]{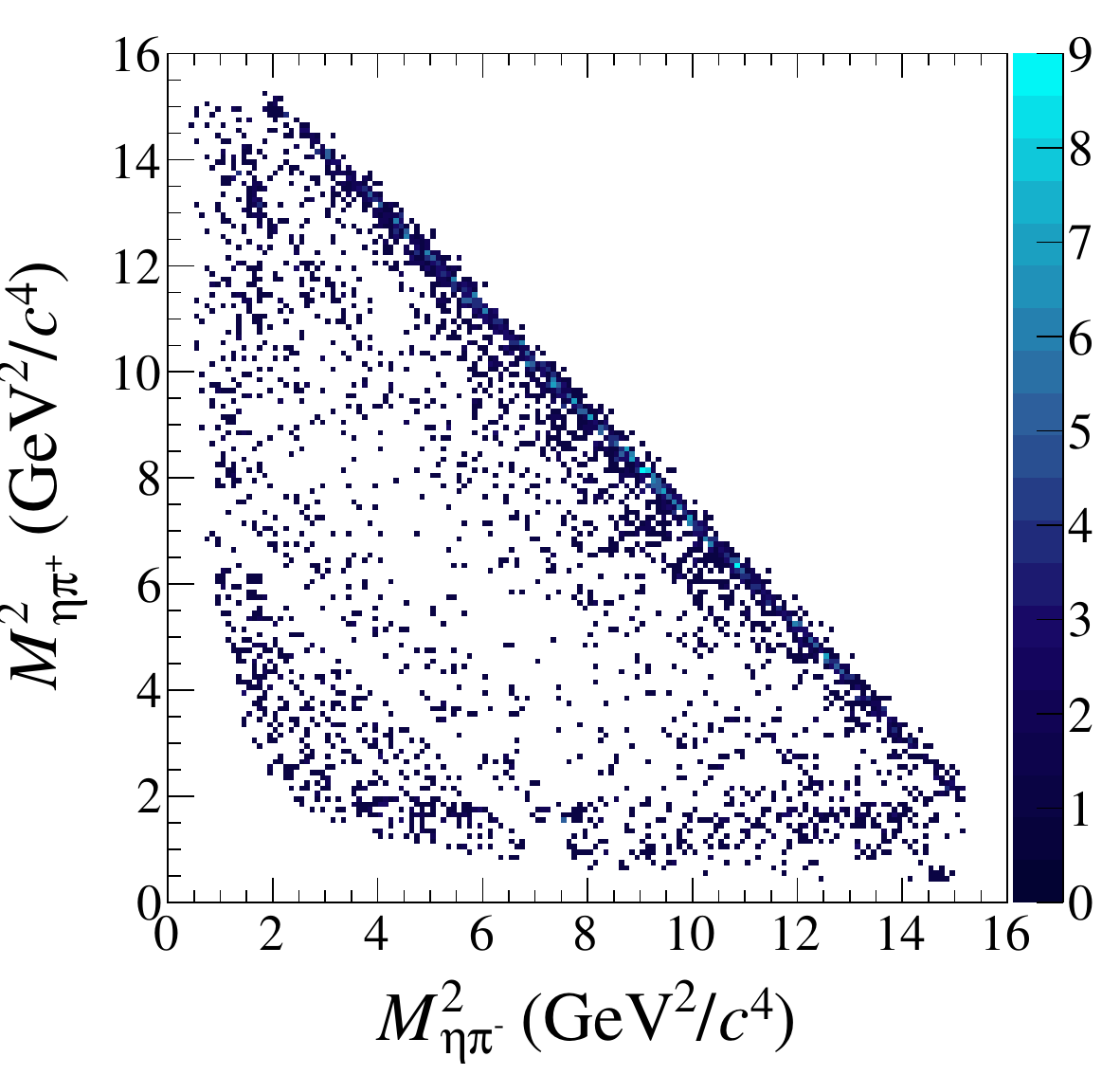}
\caption{(left) Two-dimensional distributions of
  $M_{\gamma\gamma}$ versus $M_{\pi^{+}\pi^{-}}$ for the candidate
  events for all energy points, and (right) Dalitz plot in the $\eta$
  signal range with the selected events for the energy point at
  $\sqrt{s}=4.180$ GeV. The diagonal band without events is due to the
  $J/\psi$-related event veto.}
\label{scatter-dalitz}
\end{figure}

With the above selection criteria, there are significant enhancements
close to the $\eta$ and $\rho^{0}$ nominal masses in the two
dimensional distribution of the invariant mass of
$\gamma\gamma~(M_{\gamma\gamma})$ and $\pp~(M_{\pp})$, as can be seen
in Figure~\ref{scatter-dalitz} (left). The $\eta$ signal region is
defined as $0.513<M_{\gamma\gamma}<0.581$ GeV/$c^{2}$, and the
lower and upper side-band regions are defined as
$0.309<M_{\gamma\gamma}<0.445$ GeV/$c^{2}$ and
$0.649<M_{\gamma\gamma}<0.785$ GeV/$c^{2}$,
respectively. Figure~\ref{scatter-dalitz} (right) shows the Dalitz
plot of $M^2_{\eta\pi^+}$ versus $M^2_{\eta\pi^-}$ of events in the
$\eta$ signal region at $\sqrt{s} = 4.180$~GeV. A clear $\rho^{0}$ band
is seen, and the horizontal and vertical bands around 1.75~GeV$^{2}/c^{4}$
correspond to $\ee\to a_2(1302)^{\pm}\pi^{\mp}$.

\section{Signal yields}\label{section:yield}
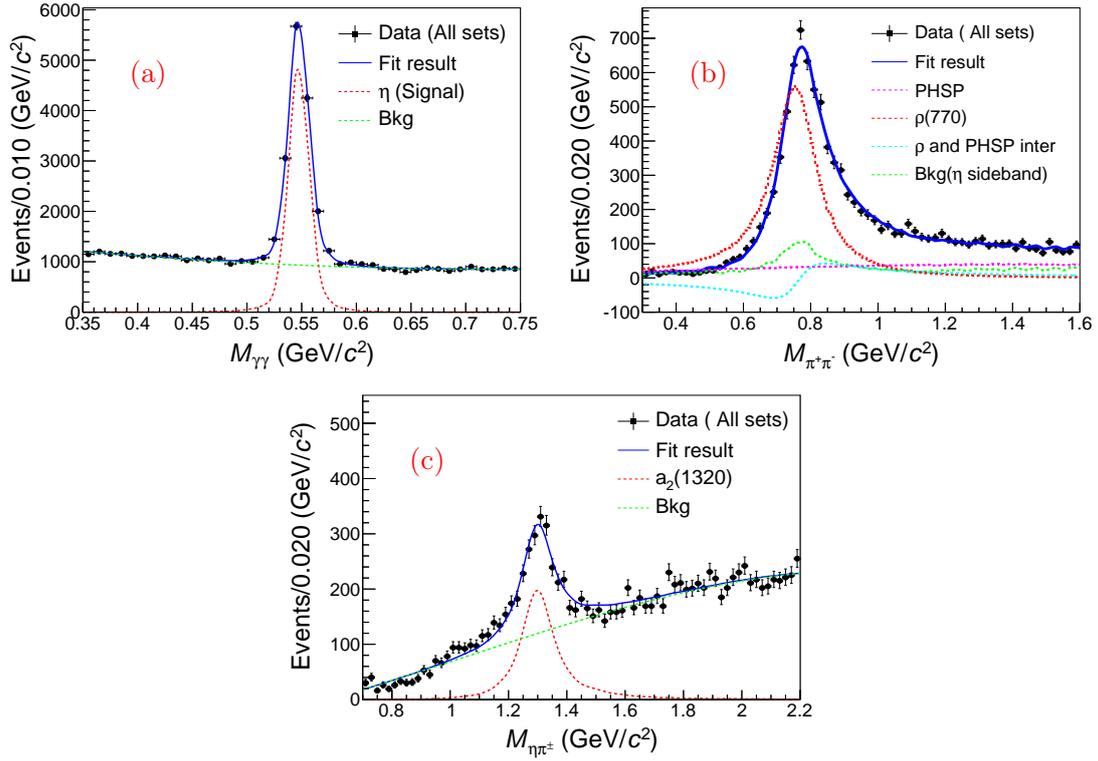
\begin{figure}[htbp] \centering
\includegraphics[width=0.47\textwidth]{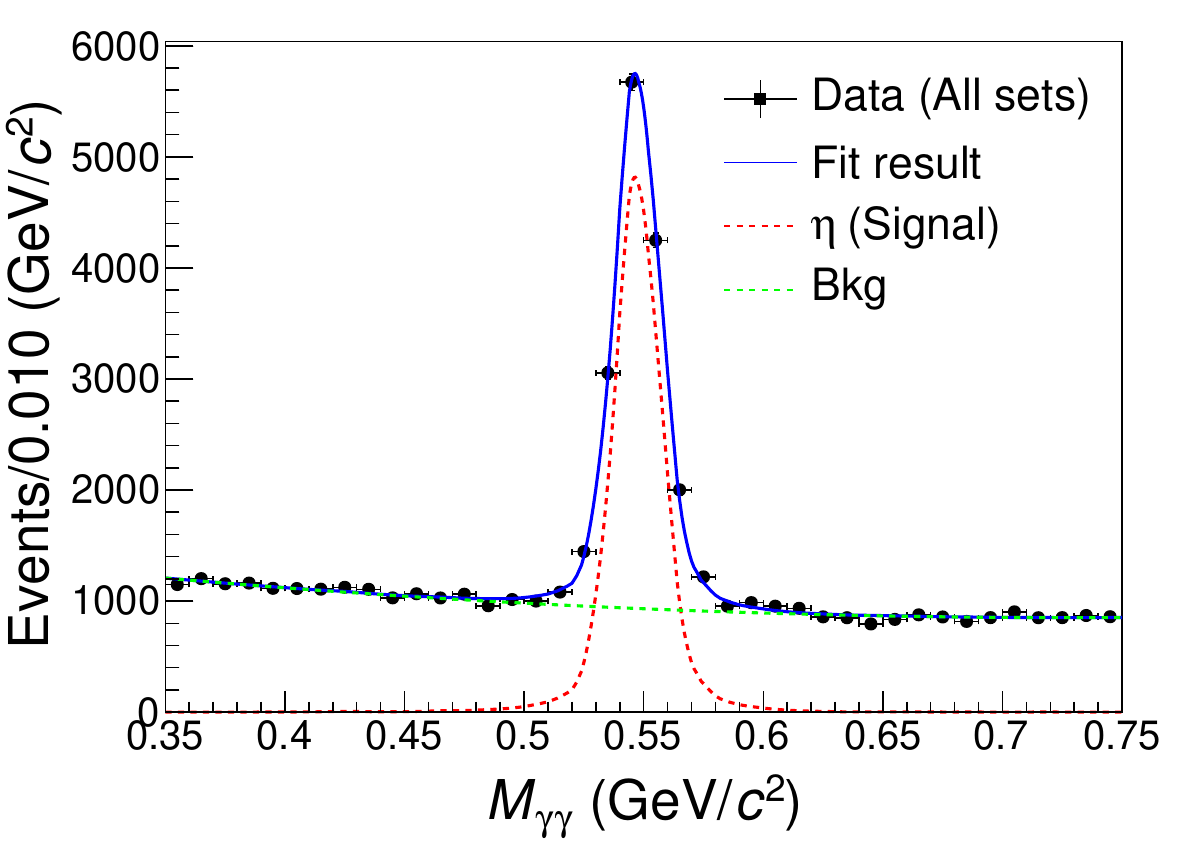}
\begin{picture}(0,0)
\put(-160,110){\color{red}{(a)}}
\end{picture}
\includegraphics[width=0.47\textwidth]{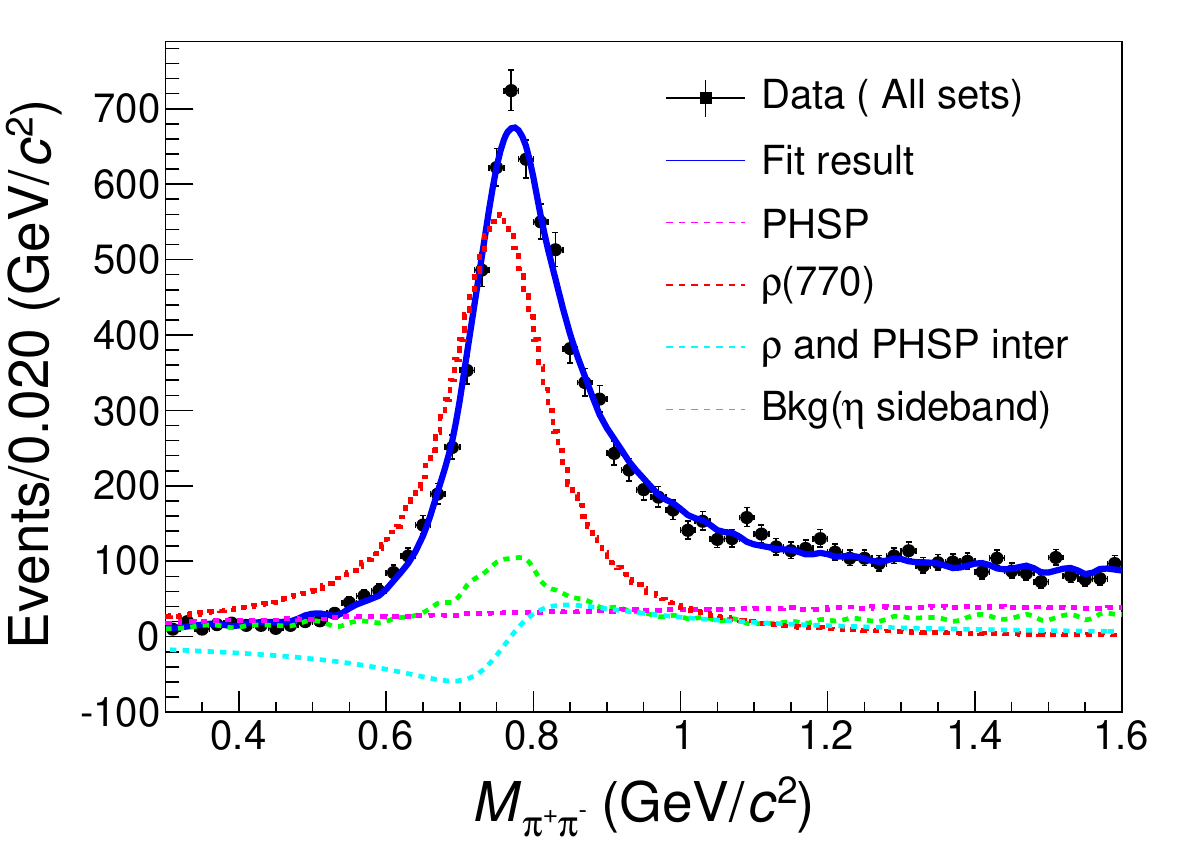}
\begin{picture}(0,0)
\put(-160,110){\color{red}{(b)}}
\end{picture}
\includegraphics[width=0.47\textwidth]{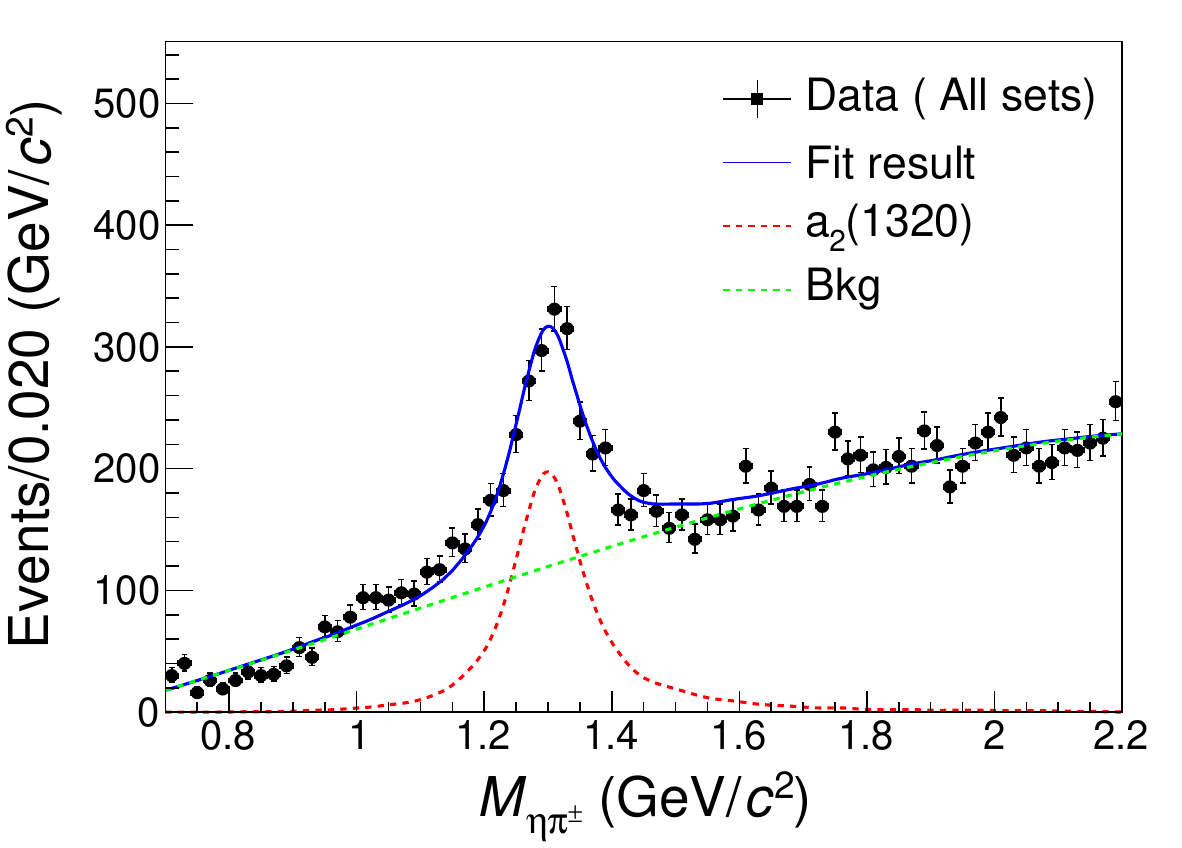}
\begin{picture}(0,0)
\put(-160,110){\color{red}{(c)}}
\end{picture}

\caption{Fit result of (a) $M_{\gamma\gamma}$, (b) $M_{\pp}$ and (c)
  $M_{\eta\pi^{\pm}}$ with the sum of all energy points. The points
  with error bars are data, the blue solid lines represent the total
  fit, the red solid and dotted lines represent the signal components,
  the green dotted lines represent the background, the green solid
  line represent the PHSP components, the cyan solid line represent
  the interference between $\rho(770)$ and PHSP process and the pink
  dotted line represent the non-$\eta$ background.}
\label{fitresult}
\end{figure}

The $\eta$ signal yields are obtained with unbinned likelihood fits to
the $M_{\gamma\gamma}$ spectra. The signal function is described as a
MC-simulated shape convolved with a Gaussian function to account for
the difference of the detector resolutions between data and MC
simulation, while the background function is described by a
second-order Chebyshev polynomial. Figure~\ref{fitresult} (a) shows
the fit result for all energy points combined.

The $\eta\rho^{0}$ signal yields are obtained with a simultaneous
unbinned likelihood fit to the $M_{\pi^{+}\pi^{-}}$ spectra of the
events in the $\eta$ signal region at all energy points. The $\rho^{0}$
resonance is parameterized by a Breit-Wigner (BW) propagator using the
Gounaris-Sakurai (GS) model~\cite{GSmodel}. The parameterized
propagator function is expressed as:
\begin{equation} \label{eq1}
{\rm{BW}^{GS}}(m) = \frac{1+d(M)\Gamma/M}{M^{2} - m + f(m,M,\Gamma) - iM\Gamma(m,M,\Gamma)},
\end{equation}
with
\begin{equation} \label{eq2}
\begin{footnotesize}
\begin{aligned}
&\Gamma(m,M,\Gamma) = \Gamma\frac{m}{M^{2}}\left(\frac{\beta_{\pi}(m)}{\beta_{\pi}(M^{2})}\right)^{3},\\
&d(M) = \frac{3}{\pi} \frac{M_{\pi}^{2}}{k^{2}(M^{2})} \ln\left(\frac{M+2k(M^{2})}{2M_{\pi}}\right) + \frac{M}{2\pi k(M^{2})} - \frac{M^{2}_{\pi}M}{\pi k^{3}(M^{2})}, \\
&f(m,M,\Gamma) = \frac{\Gamma M^{2}}{k^{3}(M^{2})}[k^{2}(m)(h(m)-h(M^{2}))+(M^{2}-m)k^{2}(M^{2})h'(M^{2})],
\end{aligned}
\end{footnotesize}
\end{equation}
where
\begin{equation} \label{eq3}
\begin{footnotesize}
\begin{aligned}
&\beta_{\pi}(m) = \sqrt{1-4M^{2}_{\pi}/m},\\
&k(m)= \frac{1}{2}\sqrt{m}\beta_{\pi}(m),\\
&h(m)= \frac{2}{\pi}\frac{k(m)}{\sqrt{m}} \ln\left(\frac{\sqrt{m}+2k(m)}{2M_{\pi}}\right),
\end{aligned}
\end{footnotesize}
\end{equation}
and $h'(m)$ is the derivative of $h(m)$, $m$ is the square of the
invariant mass of $\pp$, $M_{\pi}$ is the invariant mass of the $\pi$
meson, and $M$ and $\Gamma$ are the mass and width of the $\rho$. The
signal function is described by a coherent probability density
function (PDF):
\begin{equation} \label{eq:sigmaobs}
 {\rm{PDF}}(m) = |{\rm {BW}^{GS}}(m) + {\it{A}} \times {\rm Poly_{PHSP}} \times e^{i\varphi} |^{2},
\end{equation}
where $\varphi$ is the relative phase between the $\rho$ and PHSP
amplitudes which describe the non-$\rho$ mode parameterized with a polynomial, and the parameter $A$ is
the normalization factor. The parameters of the signal
function are left free in the fit. The non-$\eta$ background shape is
obtained by the normalized $\eta$ side-bands summed over all energies. The
number of background events is fixed to $f\cdot N_{\rm sb}$, where $f
= 0.25$ is the scale factor since the non-$\eta$ background shape is a
linear one and the side-band region is two times wider than the signal
region. The parameter $N_{\rm sb}$ is the number of side-band
background events at each c.m.\ energy point. The fits at each point
share the parameters of the signal function
(Eq.~\eqref{eq:sigmaobs}). Figure~\ref{fitresult} (b) shows the fit
result for the sum of all energy points.

The $a_{2}(1320)$ signal yield is obtained by a binned likelihood fit
to the invariant mass of $\eta\pi^{\pm}~(M_{\eta\pi^{\pm}})$ spectrum
summed over all energy points. The signal function is also described
by the MC-simulated shape convolved with a Gaussian function, and the
background function is described by a third-order Chebyshev
polynomial. Figure~\ref{fitresult} (c) shows the fit result.

Numerical results for the fits for events summed over all energy
points can be found in Table~\ref{resultratio}. The number of
$a_{2}^{\pm}(1320)\pi^{\mp}$ events is obtained by the fit to the
$M_{\eta \pi^\pm}$ distribution, the number $\eta\rho^{0} +
\rm{interference}$ events is obtained by the fit to the
$M_{\pi^+\pi^-}$ distribution, and number of
$\eta\pi^{+}\pi^{-}$ (total) events is obtained by the fit to the
$M_{\gamma \gamma}$ distribution. The number of events for the 3-body
non-resonant process is given by $N(\eta\pi^{+}\pi^{-} ~\mbox{(3-body~non-resonant)})$ $ = N(\eta\pi^{+}\pi^{-} ~ \rm{(total)}) - N(\eta\rho^{0} +
\rm{interference}) -N(a_{2}^{\pm}(1320)\pi^{\mp})$.

\begin{table*}[htbp]
\begin{center}
\caption{The numerical results of each component summed over all
  energy points. $\eta\pi^{+}\pi^{-}$(total) is obtained by a fit to
  the $M_{\gamma \gamma}$.}
\label{resultratio}
\begin{tabular}{|l|c|c|}\hline
Source                                                       &$N_{\rm{obs}}$          &$\epsilon$        \\\hline
$a_{2}^{\pm}(1320)\pi^{\mp}$                                 & 1729.7  $\pm$ 85.8     &0.2638      \\
$\eta\rho^{0} + \rm{interference}$                           & 6166.9  $\pm$ 75.7     &0.2488      \\
$\eta\pi^{+}\pi^{-}$(3-body non-resonant)                    & 4572.4  $\pm$ 181.5    &0.2718       \\
$\eta\pi^{+}\pi^{-}$(total)                                  & 12469.0 $\pm$ 140.9    &0.2593        \\
\hline
\end{tabular}
\end{center}
\end{table*}
\section{\boldmath Cross section calculation}
 The Born cross section at each energy point is calculated as:
\begin{equation} \label{eqobs}
\sigma^{\rm Born} = \frac {N_{\rm obs}}{\mathcal{L} \times \epsilon \times (1 + \delta^{\gamma}) \times \frac{1}{|1-\Pi|^{2}} \times \mathcal{B}(\eta\to\gamma\gamma)},
\end{equation}
where $N_{\rm obs}$ is the number of observed signal events,
$\mathcal{L}$ is the integrated luminosity, $\epsilon$ is the
detection efficiency, and $1 + \delta^{\gamma}$ and
$\frac{1}{|1-\Pi|^{2}}$ are ISR and vacuum polarization
(VP) factors, respectively. To obtain $1 + \delta^{\gamma}$ and
$\frac{1}{|1-\Pi|^{2}}$, we use an energy-dependent power function
$a/s^{n}$ as the initial input of the Born cross section, and the final
one is obtained by iterating several times until the difference of
$\epsilon\cdot(1 + \delta^{\gamma})$ between the last two iterations
is less than 1\%. The relevant numbers related to the Born cross
section measurement for $e^{+}e^{-}\to\eta\pi^{+}\pi^{-}$ and its
intermediate process $e^{+}e^{-}\to\eta\rho^0$ are listed in
Tables~\ref{resultpipi1} and~\ref{resultetarho1}, respectively. For
the intermediate process $\ee\to a_2(1320)^{\pm}\pi^{\mp}$, we do not
report the measurement of its Born cross section due to the low
statistics at single energy points.
\begin{table*}
\begin{center}
\caption{ Numerical results for $e^{+}e^{-}\to\eta\pi^{+}\pi^{-}$. The first uncertainties for cross sections in the most right column are statistical uncertainties and the second ones are systematic uncertainties, while those for $N_{\rm obs}$ are statistical only.}
\label{resultpipi1}
\begin{tabular}{|c|cccccc|}
\hline
$\sqrt{s}$~(GeV)  &$\mathcal{L}_{i}$~(pb$^{-1}$) &$N_{\rm{obs}}$ &$\epsilon$~(\%) &$1 + \delta^{\gamma}$ &$\frac{1}{|1-\Pi|^{2}}$  &$\sigma^{\rm{Born}}~(\rm{pb})$  \\\hline
3.872    & ~219.2   & ~243.4 $\pm$ 19.3       & 29.2 &0.8925  &1.0505  & 10.3 $\pm$ 0.8 $\pm$ 0.6 \\
4.009    & ~482.0   & ~485.8 $\pm$ 26.9       & 25.2 &0.9429  &1.0437  & 10.3 $\pm$ 0.6 $\pm$ 0.6 \\
4.130    & ~401.5   & ~361.2 $\pm$ 23.4       & 27.7 &0.9783  &1.0525  & ~8.0 $\pm$ 0.5 $\pm$ 0.4  \\
4.160    & ~408.7   & ~304.4 $\pm$ 22.1       & 27.5 &0.9804  &1.0534  & ~6.7 $\pm$ 0.5 $\pm$ 0.4  \\
4.180    & 3194.5   & 2648.8 $\pm$ 64.5       & 25.9 &0.9814  &1.0543  & ~7.9 $\pm$ 0.2 $\pm$ 0.4   \\
4.190    & ~526.7   & ~375.5 $\pm$ 24.6       & 26.5 &0.9871  &1.0558  & ~6.6 $\pm$ 0.4 $\pm$ 0.4  \\
4.200    & ~526.0   & ~411.7 $\pm$ 25.7       & 26.5 &0.9892  &1.0565  & ~7.2 $\pm$ 0.4 $\pm$ 0.4  \\
4.210    & ~517.1   & ~405.7 $\pm$ 24.4       & 26.2 &0.9921  &1.0567  & ~7.2 $\pm$ 0.4 $\pm$ 0.4  \\
4.220    & ~514.6   & ~415.4 $\pm$ 25.4       & 26.2 &0.9950  &1.0563  & ~7.4 $\pm$ 0.5 $\pm$ 0.4  \\
4.230    & 1056.4   & ~810.7 $\pm$ 35.6       & 23.7 &0.9974  &1.0560  & ~7.8 $\pm$ 0.3 $\pm$ 0.4   \\
4.237    & ~530.3   & ~393.2 $\pm$ 24.9       & 26.2 &1.0028  &1.0554  & ~6.8 $\pm$ 0.4 $\pm$ 0.4  \\
4.246    & ~538.1   & ~385.4 $\pm$ 25.9       & 26.3 &0.9986  &1.0555  & ~6.6 $\pm$ 0.4 $\pm$ 0.4  \\
4.260    & ~828.4   & ~608.4 $\pm$ 30.8       & 23.5 &1.0049  &1.0534  & ~7.5 $\pm$ 0.4 $\pm$ 0.4  \\
4.270    & ~531.1   & ~376.9 $\pm$ 24.5       & 26.2 &1.0058  &1.0531  & ~6.5 $\pm$ 0.4 $\pm$ 0.4  \\
4.290    & ~502.4   & ~309.5 $\pm$ 22.2       & 26.6 &1.0032  &1.0526  & ~5.6 $\pm$ 0.4 $\pm$ 0.3  \\
4.315    & ~501.2   & ~319.8 $\pm$ 22.6       & 26.7 &1.0116  &1.0520  & ~5.7 $\pm$ 0.4 $\pm$ 0.3  \\
4.340    & ~505.0   & ~317.4 $\pm$ 23.0       & 26.6 &1.0171  &1.0507  & ~5.6 $\pm$ 0.4 $\pm$ 0.3  \\
4.360    & ~543.9   & ~342.4 $\pm$ 22.8       & 23.0 &1.0198  &1.0511  & ~6.5 $\pm$ 0.4 $\pm$ 0.3  \\
4.380    & ~522.7   & ~337.2 $\pm$ 23.9       & 26.5 &1.0269  &1.0513  & ~5.7 $\pm$ 0.4 $\pm$ 0.3  \\
4.400    & ~507.8   & ~298.2 $\pm$ 22.8       & 26.4 &1.0297  &1.0514  & ~5.2 $\pm$ 0.4 $\pm$ 0.3  \\
4.420    & 1043.9   & ~578.6 $\pm$ 29.1       & 26.1 &1.0357  &1.0524  & ~4.9 $\pm$ 0.2 $\pm$ 0.3   \\
4.440    & ~569.9   & ~318.9 $\pm$ 23.1       & 26.3 &1.0361  &1.0542  & ~4.9 $\pm$ 0.4 $\pm$ 0.3  \\
4.600    & ~586.9   & ~244.8 $\pm$ 19.9       & 25.1 &1.0660  &1.0546  & ~3.7 $\pm$ 0.3 $\pm$ 0.2  \\
4.620    & ~511.1   & ~202.3 $\pm$ 20.7       & 25.7 &1.0692  &1.0544  & ~3.5 $\pm$ 0.4 $\pm$ 0.2   \\
4.640    & ~541.4   & ~227.2 $\pm$ 19.4       & 25.7 &1.0693  &1.0544  & ~3.7 $\pm$ 0.3 $\pm$ 0.2   \\
4.660    & ~523.6   & ~209.5 $\pm$ 18.9       & 25.5 &1.0746  &1.0544  & ~3.5 $\pm$ 0.3 $\pm$ 0.2   \\
4.680    & 1631.7   & ~575.8 $\pm$ 31.3       & 25.5 &1.0806  &1.0544  & ~3.1 $\pm$ 0.2 $\pm$ 0.2   \\
4.700    & ~526.2   & ~192.7 $\pm$ 18.3       & 25.4 &1.0802  &1.0545  & ~3.2 $\pm$ 0.3 $\pm$ 0.2   \\
\hline
\end{tabular}
\end{center}
\end{table*}

\begin{table*}
\begin{center}
\caption{Numerical results for $e^{+}e^{-}\to\eta\rho^{0}$. The first uncertainties for cross sections in the last column are statistical uncertainties and the second ones are systematic, while those for $N_{\rm{obs}}$ are statistical only.}
\label{resultetarho1}
\begin{tabular}{|c|cccccc|}
\hline
$\sqrt{s}$~(GeV)&$\mathcal{L}_{i}$~(pb$^{-1}$) &$N_{\rm{obs}}$ &$\epsilon$~(\%) &$1 + \delta^{\gamma}$ &$\frac{1}{|1-\Pi|^{2}}$ &$\sigma^{\rm{Born}}~(\rm{pb})$  \\\hline
3.872          &\ 219.2    &\  125.3 $\pm$ 10.4  &27.9  &0.8893  &1.0505   &5.6 $\pm$ 0.5 $\pm$ 0.3\\
4.009          &\ 482.0    &\  216.2 $\pm$ 13.4  &25.2  &0.9456  &1.0437   &4.6 $\pm$ 0.3 $\pm$ 0.3\\
4.130          &\ 401.5    &\  178.2 $\pm$ 11.8  &25.4  &0.9828  &1.0525   &4.3 $\pm$ 0.3 $\pm$ 0.2\\
4.160          &\ 408.7    &\  143.5 $\pm$ 10.9  &25.4  &0.9887  &1.0534   &3.4 $\pm$ 0.3 $\pm$ 0.2\\
4.180          & 3194.5    &  1286.6 $\pm$ 29.2  &24.4  &0.9934  &1.0543   &4.0 $\pm$ 0.1 $\pm$ 0.2 \\
4.190          &\ 526.7    &\  175.0 $\pm$ 11.9  &24.8  &0.9971  &1.0558   &3.2 $\pm$ 0.2 $\pm$ 0.2\\
4.200          &\ 526.0    &\  198.3 $\pm$ 12.6  &25.0  &0.9987  &1.0565   &3.6 $\pm$ 0.2 $\pm$ 0.2\\
4.210          &\ 517.1    &\  185.4 $\pm$ 12.2  &24.5  &1.0009  &1.0567   &3.5 $\pm$ 0.2 $\pm$ 0.2\\
4.220          &\ 514.6    &\  203.7 $\pm$ 12.6  &24.1  &1.0024  &1.0563   &3.9 $\pm$ 0.2 $\pm$ 0.2\\
4.230          & 1056.4    &\  381.9 $\pm$ 17.4  &24.0  &1.0038  &1.0560   &3.6 $\pm$ 0.2 $\pm$ 0.2 \\
4.237          &\ 530.3    &\  190.1 $\pm$ 12.4  &24.3  &1.0056  &1.0554   &3.5 $\pm$ 0.2 $\pm$ 0.2\\
4.246          &\ 538.1    &\  188.6 $\pm$ 12.3  &24.4  &1.0071  &1.0555   &3.4 $\pm$ 0.2 $\pm$ 0.2\\
4.260          &\ 828.4    &\  278.6 $\pm$ 15.1  &23.5  &1.0124  &1.0534   &3.4 $\pm$ 0.2 $\pm$ 0.2\\
4.270          &\ 531.1    &\  190.6 $\pm$ 12.3  &24.1  &1.0144  &1.0531   &3.5 $\pm$ 0.2 $\pm$ 0.2\\
4.290          &\ 502.4    &\  142.5 $\pm$ 10.9  &24.0  &1.0207  &1.0526   &2.8 $\pm$ 0.2 $\pm$ 0.2\\
4.315          &\ 501.2    &\  162.6 $\pm$ 11.4  &24.3  &1.0304  &1.0520   &3.1 $\pm$ 0.2 $\pm$ 0.2\\
4.340          &\ 505.0    &\  155.4 $\pm$ 11.3  &23.9  &1.0338  &1.0507   &3.0 $\pm$ 0.2 $\pm$ 0.2\\
4.360          &\ 543.9    &\  157.4 $\pm$ 11.4  &22.8  &1.0350  &1.0511   &3.0 $\pm$ 0.2 $\pm$ 0.2\\
4.380          &\ 522.7    &\  155.0 $\pm$ 11.1  &23.7  &1.0475  &1.0513   &2.9 $\pm$ 0.2 $\pm$ 0.2\\
4.400          &\ 507.8    &\  131.9 $\pm$ 10.4  &23.7  &1.0515  &1.0514   &2.5 $\pm$ 0.2 $\pm$ 0.1\\
4.420          & 1043.9    &\  264.6 $\pm$ 14.8  &24.1  &1.0533  &1.0524   &2.4 $\pm$ 0.1 $\pm$ 0.1 \\
4.440          &\ 569.9    &\  142.2 $\pm$ 10.8  &23.7  &1.0554  &1.0542   &2.4 $\pm$ 0.2 $\pm$ 0.1\\
4.600          &\ 586.9    &\  106.5 $\pm$~~9.4  &23.3  &1.0871  &1.0546   &1.7 $\pm$ 0.2 $\pm$ 0.1\\
4.620          &\ 511.1    &~~  82.8 $\pm$~~8.2  &22.5  &1.0953  &1.0544   &1.6 $\pm$ 0.2 $\pm$ 0.1 \\
4.640          &\ 541.4    &\  108.0 $\pm$~~9.0  &22.8  &1.0955  &1.0544   &1.9 $\pm$ 0.2 $\pm$ 0.1 \\
4.660          &\ 523.6    &~~  82.7 $\pm$~~8.2  &22.5  &1.0979  &1.0544   &1.5 $\pm$ 0.2 $\pm$ 0.1 \\
4.680          & 1631.7    &~  259.1 $\pm$ 13.4  &22.8  &1.1013  &1.0544   &1.5 $\pm$ 0.1 $\pm$ 0.1 \\
4.700          &\ 526.2    &~~  72.0 $\pm$~~7.6  &22.7  &1.1033  &1.0545   &1.3 $\pm$ 0.1 $\pm$ 0.1 \\
\hline
\end{tabular}
\end{center}
\end{table*}

\section{\boldmath Systematic uncertainty}
The uncertainties in the Born cross section measurements include those
of the luminosity measurement, tracking and PID efficiency, photon
detection efficiency, intermediate state, $R$ ratio, $E/cp$ ratio,
decay depth in the $\mu$ counter, $\eta$ mass window, $\jpsi$ veto,
fit of $M_{\gamma\gamma}$ and $M_{\pp}$, kinematic fit, ISR and VP
correction and detection efficiency.

\begin{itemize}
\item $\it{Luminosity~measurement}$. The luminosity is measured using
  Bhabha events with uncertainty of 1\% at all energy
  points~\cite{Lum1,Lum2}, which is taken as the systematic
  uncertainty from the luminosity measurement.

\item $\it{Tracking~efficiency}$. The pion tracking efficiency is
  determined by using the control sample $\jpsi\to
  p\bar{p}\pi^{+}\pi^{-}$. The difference between data and the MC
  simulation tracking efficiency is 1\% per track~\cite{systrk}.

\item $\it{PID~efficiency}$. The uncertainty related to the pion PID
  efficiency is studied with the sample $\ee\to K^{+}K^{-}\pp$, and
  the average difference of the PID efficiency between data and MC
  simulation is determined to be 1\% for each charged pion, which is
  taken as the systematic uncertainty~\cite{syspid}.

\item $\it{Photon~detection~efficiency}$. The uncertainty caused by photon reconstruction is 1\% per photon, which is studied by the control sample $J/\psi\to\gamma\pi^{0}\pi^{0}$~\cite{sysphoton}.

\item $\it{Intermediate~decay}$. The uncertainty due to the branching
  fraction ${\cal B}(\eta\to\gamma\gamma)$ is 0.5\% from the
  PDG~\cite{BESIII8}.

\item $\it{R~ratio}$. The uncertainty caused by the $R=\frac{\mid
  E_{\gamma_1} - E_{\gamma_2}\mid}{p_{\eta}}$ requirement is estimated
  by changing the range by $\pm 0.05$. The larger differences with and
  without changes are taken as the corresponding uncertainties.

\item $\it{E/cp~ratio}$. The uncertainty caused by
  the $E/cp$ ratio requirement is estimated from the
  control sample $\jpsi\to\pi^0\pp$. The difference between data and
  MC simulation is found to be 3.88\%, which is taken as the systematic
  uncertainty.

\item $\it{Decay~depth~in~the~\mu~counter}$. The systematic
  uncertainty caused by the requirement on the decay depth in the
  $\mu$ counter is also estimated from the control
  sample $\jpsi\to\pi^0\pp$. The difference between data and MC
  simulation is found to be 0.34\% and is taken as the systematic
  uncertainty.

\item $\it{\eta~mass~window}$. The systematic uncertainty associated
  with the $\eta$ mass window requirement is estimated by changing the
  mass window range by $\pm1\sigma$, where $\sigma$ is the $\eta$ mass
  resolution, the larger difference with and without change is taken
  as the systematic uncertainty.

\item $\it{\jpsi~veto}$. The systematic uncertainty from the
  $\jpsi$-related background veto is estimated by changing the $\jpsi$
  mass window from [3.05,3.15] GeV/$c^{2}$ to [3.06,3.14] and
  [3.04,3.16] GeV/$c^{2}$, and the larger difference with and without the
  change is taken as the systematic uncertainty.

\item $\it{Fit~of~M_{\gamma\gamma}}$. The systematic uncertainties
  associated with the fit of the $M_{\gamma\gamma}$ spectrum are
  caused by the background shape and fit range. They are estimated by
  changing the order of the Chebychev polynomial function from second
  to third and changing the fit range from [0.35, 0.75] GeV/$c^{2}$ to
  [0.40, 0.80] and [0.30, 0.70] GeV/$c^{2}$. The resulting differences
  with and without change are taken as the systematic uncertainties.

\item $\it{Fit~of~M_{\pp}}$. The systematic uncertainties associated
  with the fit of the $M_{\pp}$ spectrum come from the choice of the
  signal function, background function and fit range. They are
  estimated by:
\begin{itemize}
      \item fixing the parameters of the BW function to the values
        from the PDG;
       \item changing the order of the Chebychev polynomial function from second to third;
       \item changing the fit range from [0.30, 1.60] GeV/$c^{2}$ to [0.35, 1.65] and [0.25, 1.55] GeV/$c^{2}$.
\end{itemize}
      The resulting differences with and without change are taken as the systematic uncertainties.

\item $\it{Kinematic~fit}$. The uncertainty due to the kinematic fit requirements is estimated by correcting the helix parameters of charged tracks according to the method described in Ref.~\cite{syskine}. The difference between detection efficiencies obtained from MC samples with and without this correction is taken as the uncertainty.

\item $\it{ISR~and~VP~correction}$. As mentioned in
  Section~\ref{section:fit}, we use the energy-dependent power
  function $f(\sqrt{s}) = a/s^{n}$ to fit the line shape. The
  systematic uncertainty from the $\rm{ISR}$ and VP correction is
  estimated by varying the $n$ value by $\pm1\sigma$, where $\sigma$
  is the statistical uncertainty of the fitted $n$ value. The larger
  difference of the cross sections caused by the above changes is
  taken as the systematic uncertainty.

\item $\it{Detection~efficiency}$. The detection efficiency is
  obtained by a weighted average for the three different
  processes. The weight factors are the respective numbers of signal
  events. We randomly change the number of signal events for each
  process according to its statistical uncertainty and get new ratios
  between different processes. We mix the three processes with the new
  ratios and get new efficiencies. By repeating the above procedure,
  we obtain a group of detection efficiencies, which is almost a
  Gaussian distribution. The corresponding standard deviation is taken
  as the uncertainty caused by the detection efficiency. It is found
  that it is negligible.
\end{itemize}

Due to the limited sample size at other c.m.\ energies, the systematic
uncertainties from the event selection, mass window requirement and
background veto are taken to be the same as those at $\sqrt{s}=4.180$~GeV.
The total uncertainty in the cross section measurement is
obtained by summing the individual contributions in quadrature, and
the dominate uncertainties come from the tracking
efficiency, PID efficiency, photon efficiency and E/cp ratio
requirement. All systematic uncertainties are summarized in
Table~\ref{sys-all-etapipi} and Table~\ref{sys-all-etarho} for the
processes $e^{+}e^{-}\to\eta\pi^{+}\pi^{-}$ and
$e^{+}e^{-}\to\eta\rho$, respectively.

\begin{table*}[htbp]\small
\begin{center}
\caption{The relative systematic uncertainties~(\%) for the process $e^{+}e^{-}\to\eta\pi^{+}\pi^{-}$.}
\label{sys-all-etapipi}
\setlength{\tabcolsep}{1mm}{
\resizebox{\textwidth}{60mm}{
\begin{tabular}{|l|cccccccccccccc|}\hline
$\sqrt{s}$~(GeV)              &3.872  &4.009  &4.130  &4.160  &4.180  &4.190  &4.200  &4.210  &4.220  &4.230  &4.237  &4.246  &4.260  &4.270 \\\hline
Luminosity measurement        &1.00  &1.00  &1.00  &1.00  &1.00  &1.00  &1.00  &1.00  &1.00  &1.00  &1.00  &1.00  &1.00  &1.00 \\
Tracking efficiency           &2.00  &2.00  &2.00  &2.00  &2.00  &2.00  &2.00  &2.00  &2.00  &2.00  &2.00  &2.00  &2.00  &2.00 \\
PID efficiency                &2.00  &2.00  &2.00  &2.00  &2.00  &2.00  &2.00  &2.00  &2.00  &2.00  &2.00  &2.00  &2.00  &2.00 \\
Photon detection efficiency   &2.00  &2.00  &2.00  &2.00  &2.00  &2.00  &2.00  &2.00  &2.00  &2.00  &2.00  &2.00  &2.00  &2.00 \\
Intermediate decay            &0.50  &0.50  &0.50  &0.50  &0.50  &0.50  &0.50  &0.50  &0.50  &0.50  &0.50  &0.50  &0.50  &0.50 \\
$R$ ratio                     &0.55  &0.55  &0.55  &0.55  &0.55  &0.55  &0.55  &0.55  &0.55  &0.55  &0.55  &0.55  &0.55  &0.55 \\
E/cp  ratio                    &3.88  &3.88  &3.88  &3.88  &3.88  &3.88  &3.88  &3.88  &3.88  &3.88  &3.88  &3.88  &3.88  &3.88 \\
Decay depth in MuC            &0.34  &0.34  &0.34  &0.34  &0.34  &0.34  &0.34  &0.34  &0.34  &0.34  &0.34  &0.34  &0.34  &0.34 \\
$\jpsi$ veto                  &0.14  &0.14  &0.14  &0.14  &0.14  &0.14  &0.14  &0.14  &0.14  &0.14  &0.14  &0.14  &0.14  &0.14 \\
Fit range                     &0.21  &0.47  &0.58  &0.30  &0.13  &0.65  &0.39  &0.45  &0.56  &0.43  &0.27  &0.99  &0.21  &0.91 \\
Background shape              &0.12  &0.01  &0.01  &0.03  &0.01  &0.31  &0.19  &0.04  &0.01  &0.58  &0.42  &0.13  &0.04  &0.01 \\
Kinematic fit                 &0.01  &0.06  &0.04  &0.01  &0.01  &0.05  &0.05  &0.05  &0.01  &0.03  &0.05  &0.01  &0.03  &0.01 \\
ISR and VP correction         &0.38  &0.41  &0.05  &0.48  &0.75  &0.85  &0.72  &0.66  &0.45  &0.09  &0.17  &0.80  &0.16  &0.35 \\\hline
Sum                           &5.38  &5.40  &5.39  &5.39  &5.41  &5.48  &5.43  &5.42  &5.41  &5.41  &5.39  &5.51  &5.37  &5.45 \\\hline\hline
$\sqrt{s}$~(GeV)               &4.290  &4.315  &4.340  &4.360  &4.380  &4.400  &4.420  &4.440  &4.600  &4.620  &4.640  &4.660  &4.680  &4.700 \\\hline
Luminosity measurement         &1.00  &1.00  &1.00  &1.00  &1.00  &1.00  &1.00  &1.00  &1.00  &1.00  &1.00  &1.00  &1.00  &1.00\\
Tracking efficiency            &2.00  &2.00  &2.00  &2.00  &2.00  &2.00  &2.00  &2.00  &2.00  &2.00  &2.00  &2.00  &2.00  &2.00\\
PID efficiency                 &2.00  &2.00  &2.00  &2.00  &2.00  &2.00  &2.00  &2.00  &2.00  &2.00  &2.00  &2.00  &2.00  &2.00\\
Photon detection efficiency    &2.00  &2.00  &2.00  &2.00  &2.00  &2.00  &2.00  &2.00  &2.00  &2.00  &2.00  &2.00  &2.00  &2.00\\
Intermediate decay             &0.50  &0.50  &0.50  &0.50  &0.50  &0.50  &0.50  &0.50  &0.50  &0.50  &0.50  &0.50  &0.50  &0.50\\
$R$ ratio                      &0.55  &0.55  &0.55  &0.55  &0.55  &0.55  &0.55  &0.55  &0.55  &0.55  &0.55  &0.55  &0.55  &0.55\\
E/cp  ratio                     &3.88  &3.88  &3.88  &3.88  &3.88  &3.88  &3.88  &3.88  &3.88  &3.88  &3.88  &3.88  &3.88  &3.88\\
Decay depth in MuC             &0.34  &0.34  &0.34  &0.34  &0.34  &0.34  &0.34  &0.34  &0.34  &0.34  &0.34  &0.34  &0.34  &0.34\\
$\jpsi$ veto                   &0.14  &0.14  &0.14  &0.14  &0.14  &0.14  &0.14  &0.14  &0.14  &0.14  &0.14  &0.14  &0.14  &0.14\\
Fit range                      &0.66  &0.21  &0.54  &0.52  &0.56  &0.25  &0.61  &0.64  &0.31  &0.36  &0.05  &0.37  &0.98  &0.56\\
Background shape               &0.05  &0.01  &0.16  &0.03  &0.36  &0.25  &0.06  &0.01  &0.08  &0.44  &0.39  &0.07  &0.19  &0.02\\
Kinematic fit                  &0.06  &0.01  &0.07  &0.02  &0.09  &0.03  &0.05  &0.08  &0.01  &0.05  &0.03  &0.01  &0.01  &0.05\\
ISR and VP correction          &0.81  &0.51  &0.83  &0.23  &0.46  &0.17  &0.50  &0.46  &0.41  &0.79  &1.19  &0.75  &0.79  &0.98\\\hline
Sum                            &5.46  &5.39  &5.45  &5.39  &5.42  &5.38  &5.42  &5.42  &5.39  &5.45  &5.51  &5.43  &5.51  &5.48\\\hline
\end{tabular}}}
\end{center}
\end{table*}

\begin{table*}[htbp]\tiny
\begin{center}
\caption{The relative systematic uncertainties~(\%) for the process $e^{+}e^{-}\to\eta\rho^{0}$.}
\label{sys-all-etarho}
\setlength{\tabcolsep}{1mm}{
\resizebox{\textwidth}{60mm}{
\begin{tabular}{|l|cccccccccccccc|}\hline
$\sqrt{s}$~(GeV)                    &3.872  &4.009  &4.130  &4.160  &4.180  &4.190  &4.200  &4.210  &4.220  &4.230  &4.237  &4.246  &4.260  &4.270 \\\hline
Luminosity measurement              &1.00  &1.00  &1.00  &1.00  &1.00  &1.00  &1.00  &1.00  &1.00  &1.00  &1.00  &1.00  &1.00  &1.00\\
Tracking efficiency                 &2.00  &2.00  &2.00  &2.00  &2.00  &2.00  &2.00  &2.00  &2.00  &2.00  &2.00  &2.00  &2.00  &2.00\\
PID efficiency                      &2.00  &2.00  &2.00  &2.00  &2.00  &2.00  &2.00  &2.00  &2.00  &2.00  &2.00  &2.00  &2.00  &2.00\\
Photon detection efficiency         &2.00  &2.00  &2.00  &2.00  &2.00  &2.00  &2.00  &2.00  &2.00  &2.00  &2.00  &2.00  &2.00  &2.00\\
Intermediate decay                  &0.50  &0.50  &0.50  &0.50  &0.50  &0.50  &0.50  &0.50  &0.50  &0.50  &0.50  &0.50  &0.50  &0.50\\
$R$ ratio                           &3.88  &3.88  &3.88  &3.88  &3.88  &3.88  &3.88  &3.88  &3.88  &3.88  &3.88  &3.88  &3.88  &3.88\\
E/cp ratio                           &0.34  &0.34  &0.34  &0.34  &0.34  &0.34  &0.34  &0.34  &0.34  &0.34  &0.34  &0.34  &0.34  &0.34\\
$\eta$ Mass Window                  &0.30  &0.30  &0.30  &0.30  &0.30  &0.30  &0.30  &0.30  &0.30  &0.30  &0.30  &0.30  &0.30  &0.30\\
Fit range                           &0.71  &0.32  &0.18  &0.46  &0.03  &0.28  &0.32  &0.24  &0.32  &0.29  &0.18  &0.38  &0.20  &0.37\\
Background shape                    &0.28  &0.26  &0.26  &0.26  &0.27  &1.26  &1.26  &1.27  &1.26  &0.26  &1.26  &1.26  &0.26  &1.27\\
Signal shape                        &1.00  &1.04  &1.02  &1.01  &1.03  &1.02  &0.95  &0.97  &1.10  &0.99  &0.98  &1.03  &0.96  &0.92\\
Kinematic fit                       &0.33  &0.63  &0.91  &0.82  &0.91  &0.58  &1.00  &0.84  &1.01  &0.64  &0.87  &0.83  &0.92  &0.77\\
ISR and VP correction               &0.07  &0.07  &0.10  &0.20  &0.36  &0.15  &0.13  &0.18  &0.16  &0.35  &0.40  &0.49  &0.21  &0.48\\\hline
Sum                                 &5.50  &5.49  &5.52  &5.52  &5.53  &5.62  &5.67  &5.64  &5.70  &5.49  &5.66  &5.68  &5.52  &5.65\\\hline\hline
$\sqrt{s}$~(GeV)                    &4.290  &4.315  &4.340  &4.360  &4.380  &4.400  &4.420  &4.440  &4.600  &4.620  &4.640  &4.660  &4.680  &4.700\\\hline
Luminosity measurement              &1.00  &1.00  &1.00  &1.00  &1.00  &1.00  &1.00  &1.00  &1.00  &1.00  &1.00  &1.00  &1.00  &1.00\\
Tracking efficiency                 &2.00  &2.00  &2.00  &2.00  &2.00  &2.00  &2.00  &2.00  &2.00  &2.00  &2.00  &2.00  &2.00  &2.00\\
PID efficiency                      &2.00  &2.00  &2.00  &2.00  &2.00  &2.00  &2.00  &2.00  &2.00  &2.00  &2.00  &2.00  &2.00  &2.00\\
Photon detection efficiency         &2.00  &2.00  &2.00  &2.00  &2.00  &2.00  &2.00  &2.00  &2.00  &2.00  &2.00  &2.00  &2.00  &2.00\\
Intermediate decay                  &0.50  &0.50  &0.50  &0.50  &0.50  &0.50  &0.50  &0.50  &0.50  &0.50  &0.50  &0.50  &0.50  &0.50\\
$R$ ratio                           &3.88  &3.88  &3.88  &3.88  &3.88  &3.88  &3.88  &3.88  &3.88  &3.88  &3.88  &3.88  &3.88  &3.88\\
E/cp ratio                           &0.34  &0.34  &0.34  &0.34  &0.34  &0.34  &0.34  &0.34  &0.34  &0.34  &0.34  &0.34  &0.34  &0.34\\
$\eta$ Mass Window                  &0.30  &0.30  &0.30  &0.30  &0.30  &0.30  &0.30  &0.30  &0.30  &0.30  &0.30  &0.30  &0.30  &0.30\\
Fit range                           &0.35  &0.60  &0.73  &0.23  &0.32  &0.83  &0.40  &0.21  &0.30  &0.54  &0.33  &0.51  &0.30  &0.37\\
Background shape                    &0.26  &0.26  &0.27  &0.26  &0.28  &0.27  &0.27  &0.28  &0.27  &0.26  &0.26  &0.26  &0.27  &0.26\\
Signal shape                        &0.92  &0.86  &0.95  &1.09  &1.00  &1.01  &1.02  &1.00  &1.00  &0.90  &1.02  &0.90  &1.00  &0.82\\
Kinematic fit                       &0.59  &0.65  &0.98  &0.72  &0.59  &0.83  &0.67  &0.80  &0.57  &0.65  &0.36  &0.58  &0.54  &0.49\\
ISR and VP correction               &0.75  &0.20  &0.51  &0.35  &0.48  &0.47  &0.20  &0.53  &0.94  &0.41  &0.58  &0.51  &0.68  &0.89\\\hline
Sum                                 &5.52  &5.49  &5.59  &5.52  &5.50  &5.59  &5.50  &5.53  &5.56  &5.50  &5.49  &5.50  &5.52  &5.52\\\hline
\end{tabular}}}
\end{center}
\end{table*}

\section{\boldmath Fit to the Born cross sections}\label{section:fit}
\begin{figure}[tp] \centering
\includegraphics[width=0.47\textwidth]{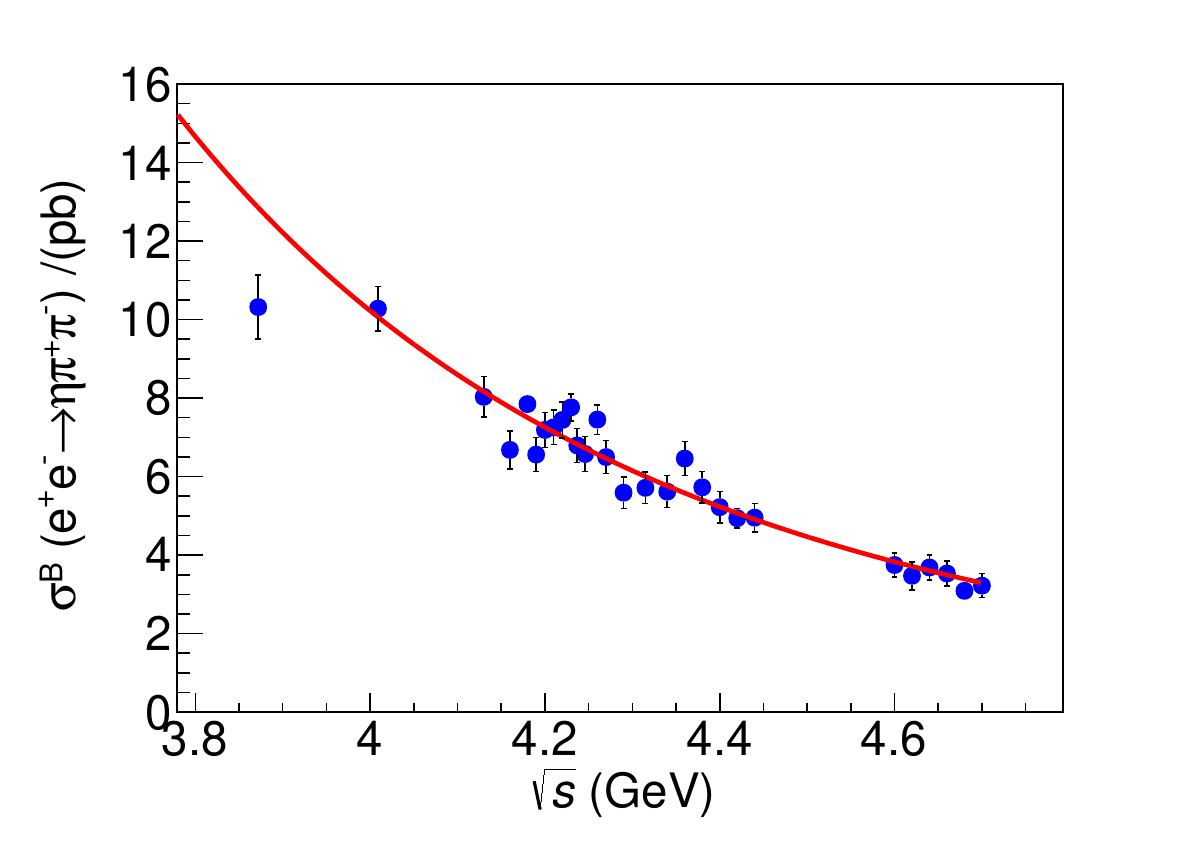}
\includegraphics[width=0.47\textwidth]{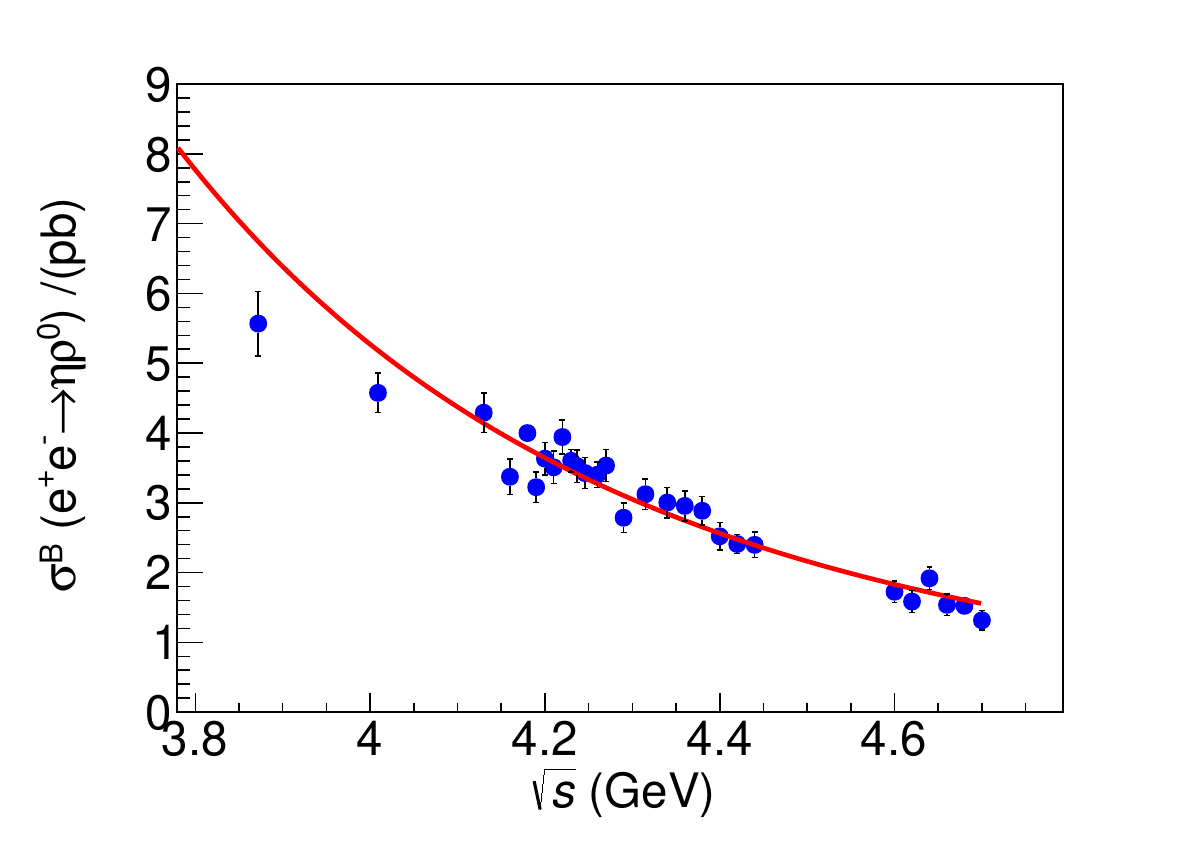}
\caption{ Fit to the Born cross section with function $a/s^{n}$ (left)
  for the $\ee\to\eta\pi^{+}\pi^{-}$ process and (right) for the
  $\ee\to\eta\rho$ process. Here, the blue dots with error bars are
  the measured Born cross sections, and the red solid lines show the
  fit results.}
\label{bornfitresult}
\end{figure}

The least-squares method is used to fit the Born cross sections under
different assumptions. In order to describe purely continuum
production, we use the energy-dependent function
\begin{equation}
	\begin{aligned}
		 f_{1}(\sqrt{s}) = a/s^{n}
		\label{eq:f1}
	\end{aligned}
\end{equation}
to fit the Born cross section which only considers the contribution
from the one photon exchange process without any resonance. The
goodness-of-fit (GOF) is $\chi^{2}/\rm{n.d.f.} = 47.0/27 = 1.7$ for
the $\eta\pp$ process and $47.5/27 = 1.8$ for the $\eta\rho$ process.
Here, $\rm{n.d.f.}$ denotes the number of degrees of freedom. The $\chi^{2}$ function is constructed as
\begin{equation}
	\begin{aligned}
  \chi^{2} = \sum\frac{(\sigma_{D_{i}} -\sigma^{\rm{fit}}_{D_{i}})^{2}} {\delta^{2}_{i}}.
		\label{eq:f4}
	\end{aligned}
\end{equation}
Here, $\sigma_{D_{i}}$ and $\sigma^{\rm{fit}}_{D_{i}}$ are the
measured and fitted Born cross sections of the $i{\rm{th}}$ energy
point, respectively, and $\delta_{i}$ is the standard deviation of the
measured cross section, which includes the statistical uncertainties only. The goodness
of the fits indicates that the data can be described by the
energy-dependent function. The fit returns $n = 3.5 \pm 0.1$ and $3.8 \pm 0.1$ for the processes $\eta\pp$ and $\eta\rho$,
respectively. The fit results are shown in
Figure~\ref{bornfitresult}. Potential contributions from the
well-established conventional charmonium states $\psi$ or charmonium-like states $Y$,
\emph{i.e.} $\psi$(4160), $Y$(4230), $Y$(4360),
$\psi$(4415), and $Y$(4660), are investigated by using
the coherent sum of the continuum (Eq.~\eqref{eq:f1})
and an additional charmonium(-like) state amplitude in the fit to the Born cross section. The fit function
can be expressed as
\begin{equation}
	\begin{aligned}
  \sigma(\sqrt{s}) = \left|f_{1}(\sqrt{s})+{\rm{BW}}(\sqrt{s}){\left(\frac{{\rm PS}(\sqrt{s})}{{\rm PS}(M)}\right)}^{1/2}e^{i\phi} \right|^{2},
		\label{eq:f3}
	\end{aligned}
\end{equation}
where the parameters $a$ and $n$ in $f_{1}(\sqrt{s})$ are fixed to those
obtained from the fit to the line shape using the function
$f_{1}(\sqrt{s})$ only. The function ${\rm{BW}}(\sqrt{s})=
\frac{\sqrt{12\pi\Gamma_{{\it{ee}}}\mathcal{B}\Gamma_{\rm
      tot}}}{s-M^{2} + iM\Gamma^{2}_{\rm tot}}$ is used to describe
charmonium(-like) states, where $M$, $\mathcal{B}$, $\Gamma_{ee}$, and
$\Gamma_{\rm{tot}}$ are the mass, branching fraction of the resonance
decays, partial width to $e^{+}e^{-}$ and total width, respectively, in which
$\Gamma_{ee}$ and $\mathcal{B}$ are left free while the other two parameters are
fixed to the values from the PDG, and $\frac{{\rm PS}(\sqrt{s})}{{\rm PS}(M)}$
is the three body phase space factor.

The statistical significances for the added components are estimated
by comparing the change of $\chi^{2}/\rm{n.d.f.}$ with and without
adding the corresponding component. Since there is interference
between the resonance and continuum process, there are two solutions
for $\Gamma_{ee}\mathcal{B}(\eta\pi^{+}\pi^{-}/\eta\rho)$ with the
same minimum value of $\chi^{2}$. Table~\ref{bwfit} lists the fit
results and the significances for the additional charmonia. The low
significances indicate that no obvious charmonium or charmonium-like
states are required to describe the measured cross section.

\begin{table*}[htbp]
\begin{center}
\caption{Results of the fits to the Born cross section. "Solution I"
  represents the constructive solution, and "Solution II" represents
  the destructive solution.}
\label{bwfit}
\setlength{\tabcolsep}{1mm}{
\begin{tabular}{|c|cccc|}\hline
Channel                                &Parameter                                     &Solution I~($10^{-4}$)   &Solution II~($10^{-1}$)       &Significance \\\hline
\multirow{5}{*}{$\eta\pi^{+}\pi^{-}$}  &$(\Gamma_{ee}\mathcal{B})^{\psi(4160)}$~(eV)  &7.4 $\pm$  \ 1.6         &\ 9.6 $\pm$ 0.1               & 1.1 $\sigma$      \\
                                       &$(\Gamma_{ee}\mathcal{B})^{Y(4230)}$~(eV)     &5.4 $\pm$  \ 3.4         &\ 7.6 $\pm$ 0.1               & 1.6 $\sigma$\\
                                       &$(\Gamma_{ee}\mathcal{B})^{Y(4360)}$~(eV)     &6.8 $\pm$  \ 2.9         & 10.9 $\pm$ 0.2               & 0.6 $\sigma$        \\
                                       &$(\Gamma_{ee}\mathcal{B})^{\psi(4415)}$~(eV)  &7.8 $\pm$  \ 1.8         &\ 6.4 $\pm$ 0.1               & 0.7 $\sigma$        \\
                                       &$(\Gamma_{ee}\mathcal{B})^{Y(4660)}$~(eV)     &5.4 $\pm$  \ 2.2         &\ 5.2 $\pm$ 0.2               & 0.5 $\sigma$        \\\hline
\multirow{5}{*}{$\eta\rho$}            &$(\Gamma_{ee}\mathcal{B})^{\psi(4160)}$~(eV)  &4.4 $\pm$  1.5           &\ 4.8 $\pm$ 0.1               & 1.0 $\sigma$     \\
                                       &$(\Gamma_{ee}\mathcal{B})^{Y(4230)}$~(eV)     &2.4 $\pm$  1.0           &\ 3.8 $\pm$ 0.1               & 1.1 $\sigma$    \\
                                       &$(\Gamma_{ee}\mathcal{B})^{Y(4360)}$~(eV)     &9.1 $\pm$  1.1           &\ 5.4 $\pm$ 0.1               & 1.6 $\sigma$    \\
                                       &$(\Gamma_{ee}\mathcal{B})^{\psi(4415)}$~(eV)  &7.5 $\pm$  2.1           &\ 3.1 $\pm$ 0.1               & 1.5 $\sigma$     \\
                                       &$(\Gamma_{ee}\mathcal{B})^{Y(4660)}$~(eV)     &3.2 $\pm$  1.1           &\ 2.5 $\pm$ 0.1               & 0.5 $\sigma$    \\
\hline
\end{tabular}}
\end{center}
\end{table*}

\section{\boldmath Summary}
The processes of $e^{+}e^{-}\to\eta\pi^{+}\pi^{-}$ and
$e^{+}e^{-}\to\eta\rho^{0}$ are studied at twenty-eight c.m.\ energies
in the energy region from 3.872 to 4.700 GeV. The Born cross sections
are obtained for all energy points. The lineshape of the Born cross
section can be well described by the empirical exponential function
Eq.~\eqref{eq:f1}. The significances for possible contributions
from $\psi$(4160), $Y$(4230), $Y$(4360), $\psi$(4415) or $Y$(4660)
resonances are all less than 2$\sigma$. This implies that the charmonium
and charmonium-like states disfavor decay to $\eta\pp$ or
$\eta\rho$.

\acknowledgments

The BESIII collaboration thanks the staff of BEPCII and the IHEP computing center for their strong support. This work is supported in part by National Key R$\&$D Program of China under Contracts Nos. 2020YFA0406300, 2020YFA0406400; National Natural Science Foundation of China (NSFC) under Contracts Nos. 11805037, 11625523, 11635010, 11735014, 11822506, 11835012, 11935015, 11935016, 11935018, 11961141012, 12022510, 12025502, 12035009, 12035013, 12061131003; the Chinese Academy of Sciences (CAS) Large-Scale Scientific Facility Program; Joint Large-Scale Scientific Facility Funds of the NSFC and CAS under Contracts Nos. U1832121, U1732263, U1832207; CAS Key Research Program of Frontier Sciences under Contract No. QYZDJ-SSW-SLH040; 100 Talents Program of CAS; INPAC and Shanghai Key Laboratory for Particle Physics and Cosmology; ERC under Contract No. 758462; European Union Horizon 2020 research and innovation programme under Contract No. Marie Sklodowska-Curie grant agreement No 894790; German Research Foundation DFG under Contracts Nos. 443159800, Collaborative Research Center CRC 1044, FOR 2359, GRK 214; Istituto Nazionale di Fisica Nucleare, Italy; Ministry of Development of Turkey under Contract No. DPT2006K-120470; National Science and Technology fund; Olle Engkvist Foundation under Contract No. 200-0605; STFC (United Kingdom); The Knut and Alice Wallenberg Foundation (Sweden) under Contract No. 2016.0157; The Royal Society, UK under Contracts Nos. DH140054, DH160214; The Swedish Research Council; U. S. Department of Energy under Contracts Nos. DE-FG02-05ER41374, DE-SC-0012069.

\end{document}